\def\simlt{\lower.5ex\hbox{$\; \buildrel < \over \sim \;$}}
\def\simgt{\lower.5ex\hbox{$\; \buildrel > \over \sim \;$}}
\def\focas{\sc focas~\rm}
\def\focasp{\sc focas\rm}
\def\iraf{\sc iraf~\rm}
\def\artdata{\sc artdata~\rm}

\documentstyle[11pt,aaspp4]{article}

\tighten

\journalid{}{}
\articleid{11}{14}

\begin{document}

\title{Colors and $K$-Band Counts of Extremely Faint Field Galaxies
\footnote{
Based, in part, on observations obtained at the W. M. Keck
Observatory, which is operated jointly by the California Institute of
Technology and the University of California}
\footnote{
Based, in part, on observations obtained at the European Southern
Observatory.}
}

\author{Leonidas A. Moustakas, Marc Davis, James R. Graham, Joseph Silk}
\affil{Astronomy Department, University of California, Berkeley, CA  94720}
\affil{e-mail: lmoustakas, mdavis, jgraham, jsilk @astro.berkeley.edu}

\author{Bruce A. Peterson}
\affil{Mt. Stromlo and Siding Spring Observatories, Institute of Advanced 
Studies,\\
The Australian National University, Private Bag, Weston Creek, \\
A.C.T. 2611, Australia}
\affil{e-mail: peterson@mso.anu.edu.au}

\and

\author{Y. Yoshii}
\affil{Institute of Astronomy, University of Tokyo, Mitaka, 
Tokyo 181 Japan\altaffilmark{3}}
\affil{e-mail: yoshii@omega.mtk.ioa.s.u-tokyo.ac.jp}

\altaffiltext{3}{Also at Research Center for the Early Universe, School of
Science, University of Tokyo, 7-3-1 Hongo, Bunkyoku, Tokyo 113, Japan}

\vskip 24pt

\centerline {Accepted by {\it The Astrophysical Journal}, 1996 July 30}

\vskip 24pt

\begin{abstract}

We combine deep $K$-band (Keck) with $V$- and $I$-band (NTT)
observations of two ``blank'' high-Galactic latitude fields, surveying
a total of $\sim2$ square arcminutes.  The $K$-band number-magnitude
counts continue to rise above $K\approx22$~mag, reaching surface
densities of few~$\times~10^5$~degree$^{-2}$.  The slope for the
galaxy counts is approximately
d~log(N)/d~mag~degree$^{-2}$~=~$0.23\pm0.02$ over the range
$18-23$~mag.  While this slope is consistent with other recent deep
$K$-band surveys, there is a definite scatter in the normalizations by
about a factor of two.  In particular, our normalization is
$\sim2\times$ greater than the galaxy counts of Djorgovski et
al. (1995).

Optical$-$near infrared color magnitude and color-color diagrams for
all objects detected in the $V+I+K$ image are plotted and discussed in
the context of grids of Bruzual-Charlot (1993, 1995) isochrone
synthesis galaxy evolutionary models.  The colors of most of the
observed galaxies are consistent with a population drawn from a broad
redshift distribution.  A few galaxies at $K\approx19-20$ are red in
both colors ($V-I\simgt3; I-K\simgt2$, consistent with being
early-type galaxies having undergone a burst of star formation at
$z\simgt5$ and viewed at $z\sim1$.  At $K\simgt20$, we find several
($\sim8$) ``red outlier'' galaxies with $I-K\simgt4$ and
$V-I\simlt2.5$, whose colors are difficult to mimic by a single
evolving or non-evolving stellar population at any redshift unless
they either have quite low metallicity or are highly reddened.  We
compare the data against the evolutionary tracks of second-burst
ellipticals, and against a grid of models that does not constrain
galaxy ages to a particular formation redshift.  The red outliers'
surface density is several per square arcminute, which is so high that
they are probably common objects of low luminosity $L<L_*$.  Whether
these are low-metallicity, dusty dwarf galaxies, or old galaxies at
high redshift, they are curious and merit spectroscopic followup.

\end{abstract}


\newpage

\section{Introduction}
The W. M. Keck telescope has opened up unprecedented opportunities for
the study of extremely faint, distant galaxies.  Deep imaging with
Keck is a highly economical and efficient mode of probing the
properties of faint field galaxies, and can yield clues for the
evolution of galaxies and cosmology.  Studying galactic evolution by
this method is best done with ``blank'' fields which avoid obvious
influence from environmental effects (such as in clusters of
galaxies), and at high Galactic latitudes, so that local Galactic
extinction and contamination by foreground stars are both minimized.
We address the classical topic of the number-magnitude relation
(``number counts''), the differential number of galaxies per apparent
magnitude per square degree.  The number-magnitude relation in the
near infrared (in particular in the $K$-band at $2.2\mu m$) is
particularly useful, because the rest-frame spectral energy
distributions of galaxies are known, and the cosmological
K-corrections are quite small out to $z>1$ and more nearly independent
of galaxy morphological type than in optical bandpasses.

Multi-color photometry allows for a reasonably detailed comparison
with models of galactic evolution while being much cheaper to do than
spectroscopy.  Combining optical with near infrared imaging offers
several advantages.  We have combined extremely deep $K$-band imaging,
obtained with the Keck telescope, with extremely deep optical $V$- and
$I$-band imaging, obtained with the New Technology Telescope (NTT), to
study the colors of faint field galaxies.  In Section 2, we describe
the observations and the characteristics of the data, and give details
of the $K$-band data reductions.  In Section 3 we present how the
aperture corrections and the completeness corrections are determined
and applied to produce the $K$-band galaxy counts.  The color
diagrams, their features, and a grid of isochrone synthesis galaxy
evolutionary models that are presented and discussed in Section 4.
Conclusions to these data and the model considerations are drawn in
Section 5.

\section{Observations and Reductions}

\subsection{The Observations} 
The optical $V$- and $I$-band data were obtained with the NTT, a 3.5-m
active-optics telescope in La Silla on 1991 March and 1991 April.  The
limiting magnitudes of galaxies in these images are
27~mag~arcsec$^{-2}$ at 3$\sigma$ in both $V$ and $I$ in more than
$26000$~s integration in each bandpass.  The typical seeing in the
data is better than $1.0$~arcsec.  These data and the related science
are discussed by Peterson et al. (1996; hereafter P96).

Near infrared $K$-band ($2.2\mu m$) data were obtained with the Near
Infrared Camera (NIRC) on the W. M. Keck 10-m telescope, on 1994
February 23 and 24 (Moustakas et al. 1995).  The detector array used
in NIRC is an InSb (Indium-Antimonide) photovoltaic $256\times 256$
square array with $29\mu m$ square pixels (Matthews \& Soifer 1994).
The resulting pixel scale is $0.15$~arcsec~pix$^{-1}$.  The weather
was clear, with the seeing varying from $\sim0.7$~arcsec to
$\sim0.9$~arcsec by the end of the second night, due to strong winds.
Because the sky brightness is quite high in the $K$-band ($\sim
14.5$~mag~arcsec$^{-2}$), individual exposures were short.  Because of
the excellent control of the thermal background emission at Keck,
there was no signal to noise advantage in using a $K'$ ($2.1 \mu m$;
Wainscoat \& Cowie 1992) filter.  A typical NIRC frame consisted of
eight consecutive in-place exposures of $15$~s ($8\times 15$~s).

Two widely separated fields, which will be called ``Field I'' (Figures
1 and 3 [Plate 1]) and ``Field II'' (Figures 2 and 4 [Plate 2]), were
observed.  Both are ``blank'' fields at high Galactic latitude
($b_{Field~I}\approx+50^{\circ}, b_{Field~II}\approx+60^{\circ}$), and
are subsections of the optical NTT fields (P96).  Coordinates for the
field centers are given in Table 1.  Both fields were observed with a
$2\times 2$ mosaic pattern, with a small ($\sim8$~arcsec) overlap.
The centers of each quadrant in the mosaic pattern are separated by
$30$~arcsec, making the total field of view approximately
$1.1\arcmin\times1.1\arcmin$ for each field.  Between each set of
eight exposures, the telescope was moved using a non-redundant pattern
in both $(x,y)$ directions, within $8$~arcsec in either direction.
With this observing plan, objects on the sky are imaged on different
parts of the array, and the data images can be used for the
sky-subtraction and the flatfielding.

The resultant integration time (and therefore the signal to noise)
varies across each field, and must be accounted for in the analysis of
these frames.  Individual objects are detected based on a number of
contiguous pixels having intensities some factor higher than the local
sky.  Thus, for the purpose of detecting objects, an ``effective sky''
is specified as follows.  Because signal to noise is proportional to
the square root of the integration time, the final mosaics for the two
fields are each multiplied by the {\it square root} of the
corresponding exposure map images.  This operation forces the
rms-``sky'' to have the same value at all points in the mosaics,
making it fair to use a single set of criteria to search for objects
over the entire mosaic.

For Field I, the time spent on each of the four pointings was
$6000$~s, reaching approximately $23.7$~mag~arcsec$^{-2}$ at 3$\sigma$
in an aperture of 0.9 arcsec diameter. For Field II, the time spent on
two of the four pointings was $4800$~s, reaching approximately
$23.1$~mag~arcsec$^{-2}$ at 3$\sigma$ in the same aperture.  There was
not enough time in the observing run to finish all four pointings of
Field II.  This is quite apparent in Figure 4 (Plate 2), where the
signal to noise of the $K$-band image is much worse for the two
neglected pointings.  In the analysis of Section 3 we only consider
the two deeper pointings of Field II.

Guiding with NIRC is accomplished via an offset guider camera.
Unfortunately, only a single guiding star was available for each
field.  NIRC and its offset guider have a fixed relative orientation,
which can be rotated relative to the instrument's axis.  Thus it was
necessary for two of the frames in each mosaic to be rotated by a
small angle ($5.7\arcdeg$) with respect to the other two.  This
relative rotation is apparent in Figures 3 (Plate 1) and 4 (Plate 2).

Although there are no bright stellar objects in any of the pointings
to use for stacking, all exposures were guided, so the offsets as
recorded at the telescope were used to construct the final coadded
mosaics.  Individual frames were shifted by the amount specified by
the recorded offsets, down to the corresponding fractional-pixel
amount.  The resulting mosaics using the fractional-pixel shifts were
found to be rather indistinguishable from mosaics that used integer
pixel shifts.  Also, the accuracy of the offsets seems to be reliable.
To satisfy ourselves that this was the case, we experimented with a
cross-correlation technique for refining the offsets, as follows.
First, the mosaics were created as described above.  Then, each
individual image was placed in the correct location on top of the
mosaic.  A cross-correlation was calculated based on small offsets of
the individual image with respect to its original position; in all
cases, the ``refined'' offsets were practically identical to the
original (recorded) offsets.  It should be noted that for many of the
individual frames, there may not have been sufficient signal in
individual objects to calculate reliable refined offsets.  To
establish the photometric zeropoint, the UKIRT faint $JHK$ standard
star FS14 was observed (Casali \& Hawarden 1994), and the frames were
reduced in the same manner as the primary target frames.

\subsection{The $K$-band Reductions}

In the $K$-band the sky brightness varies quite rapidly, and since we
are exploring levels fainter than a part in $10^4$ of the sky
brightness, it is crucial to do the sky subtraction carefully.  The
procedure for reducing the $K$-band images is described below.  All
images are dark-subtracted and linearized (Graham 1995).  Pixels that
saturate at very low intensity levels are marked as bad pixels, and
are included in a bad-pixel mask.  We proceed with a crude sky
subtraction for each frame by creating a median-average frame of the
five (temporally) nearest frames, where each frame has been normalized
to the same level as the frame we wish to sky-subtract.  This brings
out the objects, for which object masks are then created.  Once this
procedure has been completed for all frames, it is repeated with the
following differences.  All the objects as well as all bad pixels are
ignored when creating the sky frame for each frame by averaging the
five (temporally) nearest frames.  Each of the final sky-subtracted
images are flat at the level of $\sim 4\times 10^{-5}$.  The remaining
background contains fluctuations on a variety of spatial scales, and
necessitate the use of a local background for the extraction of the
flux of each galaxy.  The entire pipeline is processed using scripts
written in the \iraf environment.

\section{$K$-band Number-Magnitude Relation}

\subsection{Photometry}

The Faint Objects Classification and Analysis System (\focasp; Tyson
\& Jarvis 1979; Tyson 1988) package is used for detecting objects and for
performing the aperture and the isophotal photometry.  An ``object''
in \focas is defined as a set of contiguous pixels, each of which
meets or exceeds a minimal signal to noise ratio specified by the
user.  Each object detected has its position, shape, and integrated
flux measured and stored.  \focas measures three types of fluxes:
fixed-aperture, isophotal, and ``total'' isophotal.  We have chosen to
use a $3.0$~arcsec diameter fixed aperture for the entire
number-magnitude relation analysis.  The \focas isophotal magnitude
corresponds to the amount of light within the isophote that is traced
by the user-specified minimum signal to noise ratio, and is the
magnitude system we have used in measuring the colors of the galaxies,
as described below.  The \focas ``total'' isophotal flux is the flux
measured in an object whose isophotal area is increased by a factor of
two; it is not used in this analysis.

The aperture correction is a constant offset in magnitudes that
corrects for the missing flux of an object more extended than the
limiting (fixed) aperture.  For a given aperture size, the aperture
correction can depend on the intensity-profile an object follows, and
the distance as well.

Redshift information for each galaxy would be required to use metric
apertures for determining the number-magnitude relation, but this is
not possible at the flux levels reported here.  It is instructive to
consider the systematic effects of using aperture magnitudes when
metric magnitudes are not available.  Galaxy surface brightness
profiles can be parameterized by $$L(<r)\propto r^{\alpha}$$ (Gunn \&
Oke 1975).  For field galaxies, a reasonable value is
$\alpha\approx0.4$ (Mobasher et al. 1993; Glazebrook et al. 1995b).
Based on this relation, the amount of light in a $3.0$~arcsec diameter
aperture compared to a $20h^{-1}$~kpc
($h=H_0/$100~km~s$^{-1}$Mpc$^{-1}$) diameter metric aperture as a
function of redshift is shown in Figure 5.  For the middle of our
magnitude range ($K\approx20$~mag), the median redshift for
nonevolving galaxies in a Euclidean universe is $z\approx0.8$.  Near
that redshift, Figure 5 shows that there is an approximately constant
systematic offset of $\simlt0.20$~mag.  Because the offset is constant
within $0.1$~mag between redshifts of $0.5 < z < 1.0$, using
sufficiently large fixed-apertures for calculating magnitudes should
not introduce any strong systematic trends in the galaxies' calculated
magnitudes compared to their true magnitudes.

\subsection{Simulations and Corrections}

There are observational limitations in the data, especially near the
detection limits.  Extensive Monte-Carlo simulations were carried out
to characterize the data and to determine the best methhod of
analysis and presentation.  
For the entire analysis, the \iraf \artdata package was
used to create simulated data sets, which were then analyzed exactly
as the real data were, using the \focas programs.  

We experimented with various proportions of elliptical and spiral
galaxies in the simulations.  These galaxies were generated using the
\artdata routines, assuming no internal extinction, and a redshift
distribution that corresponded to the apparent magnitude distribution
for each iteration of the simulation.  The data were placed in a
``noise-only'' image, which simulated our real mosaic.  To create the
``noise-only'' image, a set of frames was created, each one being
equivalent to a typical NIRC exposure from our set of observations,
except with no objects in it.  The noise-frames were then mosaiced
exactly as the real data were, thus creating a mosaic equivalent to
each Field with no objects, with all the effective signal to noise
characteristics of the real mosaics.

For each half-magnitude bin over the entire range of magnitudes that
we span, five galaxies of random integrated magnitudes within that
magnitude bin were placed in random locations in the ``noise-only''
mosaic.  This was then re-analyzed normally using the \focas programs,
and the number and properties of the recovered galaxies was recorded.
For each half-magnitude bin, this procedure was repeated one hundred
times.  These simulations give us enough information about the
behavior of our mosaics to establish both a reasonable aperture
correction and the recovery efficiency at each magnitude bin for
different proportions of galaxy types, and to calculate the corrected
number-magnitude relation.  The recovery efficiency vector contains
the fraction of galaxies recovered as a function of the center of the
magnitude bin.

The simulations for different ratios of galaxy types give slightly
different aperture offsets, all within $\sim0.1$~mag of $-0.30$~mag.
The scatter about the offset does not depend strongly on the assumed
ratios.  The completeness matrix, described below, reflects exactly
where all the simulated galaxies go.  In the end, however, the
differences in the number-counts bins and slope are not very sensitive
to the final result.  We did not experiment with the  inclusion of very
diffuse or low surface brightness galaxies.  For each of our
simulation sets, we reach a $\sim50$\% recovery rate of the simulated
galaxies at $K\approx22.5$~mag with any $K<24$.  The insert in Figure
7 plots the fraction of galaxies recovered as a function of the center
of the magnitude bin for each of the galaxy type proportions and
offsets that we calculated in the simulations.  The final corrections
that were applied to the data assume a $40/60$ ratio of
ellipticals and spirals.  Figure 6 shows the output of this
simulation.  Based on this output, we shall use a constant aperture correction 
of $-0.30$~mag.

The ``completeness matrix,'' $C_{mn}$, contains all the information
about each set of simulations, and can help estimate the ``bin
jumping'' that will happen when a galaxy's magnitude is not recovered
within the magnitude range that it started out with.  If we denote the
observed number-magnitude vector as $M_{obs}$ and the true
number-magnitude vector as $M_{true}$, then $$C_{mn}\times M_{true} =
M_{obs}.$$ By inverting the completeness matrix, it is possible to
recover the true number-magnitude vector $M_{true} = C_{mn}^{-1}\times
M_{obs}.$ This approach is used routinely in studies of globular
cluster luminosity functions (e.g. Drukier et al. 1993; and Drukier et
al. 1988).  The completeness matrix for Field I is given in Table 4.

\subsection{Number-Magnitude Results}

Once the aperture offsets and the completeness corrections have been
applied, the data are binned by whole-magnitude intervals.  The galaxy
counts are presented in Table 5, and plotted in Figure 7.  The error
bars are $1\sigma$ uncertainties based on the Poisson statistics
(Gehrels 1986) of the raw counts.  For comparison, we have plotted the
results from several other deep $K$-band surveys, the deepest of which
is by Djorgovski et al. (1995), which reaches about $0.5$~mag deeper
than our data.  For the range $20\le K \le 22$, the Djorgovski et
al. numbers are below our results, which seem to be more consistent
with the other observations.  This may be due to physical variations
in the counts between fields, or at least in part to systematic
differences introduced in the data analysis or the corrections.  The
uncertainties in our photometric zeropoints are all smaller than approximately
a $-0.1$~mag, which is far too small a shift to account for
the variations that are seen.

Over the interval $18\le K \le 23$, the logarithmic slope of our
completeness-corrected galaxy counts is $\approx0.23\pm0.02$,
consistent with the other deep $K$-band surveys.  Because of the large
uncertainty in the $K=23$~mag bin, we cannot confidently confirm that
the counts continue to rise.  As a guide to the interpretation of the
counts, we have plotted models from Yoshii \& Peterson (1995) in
Figure 7.  ``Model 1'' uses $(\Omega_0,\lambda_0,h_0)$ = (1.0, 0.0,
0.5), and ``model 2'' has (0.2,0.0,0.6).  The open models (``model
2'') make a more reasonable fit to the count data alone than the
$\Omega=1$ models.  No steadfast conclusions should be drawn from this
comparison between data and models since any particular model should
simultaneously explain the counts in all bandpasses, and the redshift
distributions of all the faint galaxies.

\section{Colors and Galaxy Evolution}

\subsection{Background}

The $K$-band Keck fields were chosen because of the extant extremely
deep optical data of P96).  We combine the optical and near infrared
photometry of these extremely faint galaxies to make color-magnitude
and color-color diagrams, and compare the data with galactic evolution
models, which will be discussed in the next Section.

To obtain the objects' colors, we combine the $V$, $I$ with the $K$
mosaic (which has been multiplied by the square root of its exposure
map; see Section 2.1) by normalizing all three to have the same rms
sky-noise value and then summing them.  We call this combined $V+I+K$
image the ``$VIK$'' image.  \focas is used to detect objects and to
generate isophotes, as described in Section 3.1.  The catalog and
isophote boundaries thus created were then applied to each of the
$V,I,$ and $K$-band images individually.  By this method the
object-detection and the isophote determination is based on the
combined image and the isophotes are not biased towards any one band.
The object catalogs for both Fields are listed in Tables 2 and 3.
Because aperture corrections are not well-defined for isophotal
magnitudes, {\it no} corrections are applied to these data as they are
plotted in the color-magnitude and color-color diagrams in Figure 8
and 9.  Similar diagrams using aperture magnitudes are reasonably
comparable.  In the color-color diagram, the sizes of the points scale
with $K$-band magnitude as shown in the legend. The diagonal line in
Figure 8 is the $3\sigma$ flux limit of the $I$-band data
($I_{lim}\approx27.0$~mag).

\subsection{Models}

Although there remain significant uncertainties in galaxy evolutionary
codes (Charlot, Worthey, \& Bressan 1996), they are very powerful for
developing a broad understanding of the nature of galaxies for which
there is relatively limited photometric or spectroscopic information.
With at least three bandpasses it is possible to compare models more
directly with the data because the zeropoint normalization due to an
assumed galaxy mass cancels out.  Especially with the near infrared
data, we can directly probe the underlying old population of stars in
each galaxy.  Thus, most of the emphasis in the following section is
on the optical$-$near infrared color-color diagram (Figure 9).  The
apparent magnitudes for the models expressed in the color-magnitude
diagram (Figure 8) are normalized to a total mass for each galaxy of
$M=10^{11}M_{\odot}$.

We work primarily with a grid of Bruzual-Charlot (1993, 1995) models
spanning a large range of star formation rate (SFR) timescales viewed
at redshifts in the range $0 \simlt z \simlt 7$.  All models are for
solar metallicity, following a Scalo IMF.  The star formation
timescales used are $\tau=(0.01, 0.1, 1.0, 5.0, 10.0, 30.0)$~Gyr.  One
constant-rate star-formation model was also included, at
$6.7$~M$_{\odot}$~yr$^{-1}$.  No reddening has been included in the
models, but the reddening vector for $E(B-V)=1.00$ (based on Savage \&
Mathis 1979) is shown in the color-color diagram (Figure 9).  A
realistic inclusion of reddening to this problem would involve a more
complicated analysis than simply applying the effect of a simple
dustscreen between the galaxy and the observer.  The assumed cosmology
is $H_0 = 50$~km~s$^{-1}$~Mpc$^{-1}$ and $q_0=0.5$.

These grids of models have been examined two different ways.  First,
an explicit assumption is made for the epoch of galaxy formation.  The
models overplotted in Figure 9 assume a uniform epoch of star
formation at $z_{form} = 5.0$.  Assuming a greater formation redshift
(e.g. $z_{form} = 7.0$) does not change the results significantly,
especially considering the photometric uncertainties in the data
(typical uncertainties in the colors are shown at the lower-right
corner of Figure 9).  Using a lower $q_0$ does not change the models
significantly.  Only four of the seven evolving models are plotted in
Figures 8 and 9.  The heavy-solid line traces a non-evolving
elliptical galaxy as a function of redshift, i.e. pure K-correction;
the light-solid line is for an exponentially declining SFR with
timescale $0.1$~Gyr; the short-dashed line is for an exponentially
declining SFR with timescale $5.0$~Gyr; and the long-dash line is for
a constant SFR, equal to $6.7$~M$_{\odot}$yr$^{-1}$.

The second approach to model building decouples the age constraint
from the viewing redshift, so that galaxies are allowed to have formed
at any time, with the only restriction being a requirement that a
galaxy's age be less than the age of the universe at that epoch.
Reddening will generally move the model points toward the upper right
in color-color space (shown in Figure 9), while decreasing the model
metallicity will primarily move the color-color point downward because
of increased temperatures on the giant branch and changes in the main
sequence lifetime.
Combinations of all three of
these influences can populate much of the color-color space, but many
of the combinations make little physical sense.  While most galaxies
have optical$-$near infrared colors that are consistent with any one
of a large number of model parameter combinations, there are some
extremely faint galaxies upon whose nature one may productively
speculate.

\subsection{Features in the Color Diagrams}

Most of the galaxies have colors that are consistent with a population
drawn from a variety of redshifts and galaxy types, as is evident from
Figures 8 and 9.  This is true even with the models plotted which
assume a single formation redshift, $z_{form}=5$, and in the context
of the grids of models the range of possibilities is even greater.
However, many of the brighter galaxies in this sample
($K\approx19-20$) are red in both $V-I\simgt2$ and $I-K\simgt3$
(Figure 9); the most straightforward interpretation is that they are
early-type galaxies at $z\sim1$, though it is also possible for these
galaxies to be either very old, or extremely reddened early- {\it or}
late-type galaxies at redshifts of only a few tenths.  Either
multi-band photometry or, preferably, spectroscopy is needed to
distinguish between the possibilities.  Identifying the nature of some
of these very red galaxies may have significant implications for
setting an independent constraint on the age of the universe if they
are old galaxies at moderately high redshifts.  Recently, spectroscopy
of very red, faint galaxies (e.g. 53W091, $K=18.8$~mag, $R=24.8$~mag;
see Dunlop et al. 1996 and Spinrad et al. 1996) is revealing  examples of
very old early-type galaxies at $z\approx1.5$.

At $I-K\simgt4$, $V-I<2.5$ there are $\sim8$ galaxies that are
distinct from the galaxies described above.  For convenience, we shall
refer to them as the `red outlier' galaxies in the following
discussion.  Similar outlier objects have turned up in all deep
optical/near infrared surveys (e.g. in the surveys by Cowie et
al. 1994, 1995; Glazebrook et al. 1995; and originally by Elston et
al. 1988 and Elston et al. 1989).  The galaxies in our fields are not
quite as red as the apparently-rare objects detected by Hu \& Ridgeway
(1994), which have $(I-K)\approx6.5$, but parts of the discussion that
follows may be applicable to those galaxies as well.

The colors of the red outliers do not have an obvious correspondence
with either the pure K-correction track, or any of the passive
evolution tracks of Figure 9, which assume a single formation redshift
$z_{form}=5$.  Songaila et al. (1994) argue that galaxies from their
survey with colors similar to the red outliers are consistent with
evolving populations for galaxies in the range $1<z<2$.  This is
clearly not the case for elliptical galaxies that formed at
$z_{form}=5$, which would have considerably redder optical colors.  In
an effort to explain these red outliers, we explore several
second-burst ellipticals models, the grid of BC95 models discussed in
Section 4.2, and a qualitative scenario involving dust.  Based on the
surface density and apparent $K$-band magnitudes of the red outliers,
we discuss the relative merits of each scenario, and where possible,
make predictions that will be possible to test in the near future.

The surface density of the red outliers is substantial.  Within the
$\sim2$ square arcminutes we have surveyed, we detect $\sim8$ such
galaxies.  If these galaxies are distributed to a distance of $z
\approx 1$, then their space density is of order $n \approx 0.002
h_{100}^3$ Mpc$^{-3}$, which is only a factor ten smaller than that of
$M_{K_*}$ galaxies locally.  For an Einstein-de Sitter cosmology,
these galaxies, if placed at $z\approx 1$, would have an absolute
magnitude $M_K \approx -21.5~+~5{\rm log}(h)$, which is $\sim2$
magnitudes below $M_{K_*}$ (Mobasher et al. 1993).  Thus the red
outliers could be fairly common, not overly-luminous galaxies.

The simplest extension of the single-$z_{form}$ models involves
superimposing a judiciously timed second burst of star formation on a
passively-evolving elliptical galaxy.  We experimented with such
second-burst elliptical galaxy models (Charlot 1995a, 1995b) by adding
bursts of star formation at various redshifts and for various mass
fractions.  If only small mass fractions are involved in the burst,
then it is possible to drive the $V-I$ colors dramatically blueward
while maintaining similar $I-K$ colors.  The best combination has
early-type galaxies undergoing bursts involving $\sim1\%-5\%$ of their
mass at $z\approx3$. Such objects come close to the colors of the red outliers,
if the galaxies are viewed at redshifts no less than about 2.5.  This
would imply that the red outliers, which make up less than $10\%$ of
the galaxies at $K=20-22$~mag, all experienced a burst of star
formation at the same time, and are being viewed specifically when
they have their bluest optical (rest-ultraviolet) colors.  It is
conceivable that such a universal event could have occurred during an
epoch of merging, but it would be odd that we do not see the resulting
galaxies at lower redshifts.

By considering the grids of models described earlier, the constraint
on the particular age of a galaxy at each redshift is lifted.  At
redshifts lower than $z\sim1.5$, it is not possible to produce the
colors of the red outliers, except perhaps with a combination of very
low metallicity and a substantial amount of reddening.  

Finally, we consider a qualitative scenario, where the red outliers
are normal galaxies with dusty, old populations within their galactic
nuclei.  As a simple, but hardly unique demonstration, suppose that
the outliers are in the range $0.7 < z < 2$ and are composed of a
mixture of an old partly extincted population formed at $z=5$,
superposed with a young population of equal (unextincted) flux in the
received $V$-band.  From Figure 9, we can infer that the old
population, if unextincted, would have observed colors, $V-I \approx
3$, $I-K \approx 4$, while the young population would have colors $V-I
\approx 1$, $I-K \approx 2$.  If the old population has one magnitude
of extinction in the $K$-band, assuming that the optical depth varies
as $\lambda^{-1}$, we would expect a composite galaxy with color, $I-K
\approx 4.5$, $V-I \approx 1.5$, which is consistent with the observed
colors of the outliers.  

One cannot simply argue for large dust absorption without being
concerned about the contribution to the abundance of IR ultraluminous
sources (Graham et al. 1996; Soifer et al. 1994), or to the diffuse
100$\mu$ background (e.g. Puget et al.  1996), but the model above
leads to less than 50\% of the flux being reprocessed into the far IR.
It is very likely that extinction is important in high redshift galaxies, 
especially in view of  recent claims of the detection of  a 100$\mu$ 
cosmological background radiation (Puget et al 1996).

Ground-based searches for the predicted bright phase associated with
elliptical formation have been singularly unsuccessful (e.g. Thompson
et al. 1995; Thompson \& Djorgovski 1995; Pahre \& Djorgovski 1995),
implying that ellipticals (and spheroids) formed at high redshift,
were dust-enshrouded, or both.  Theoretical arguments which appeal to
the analogy between starbursts and the formation of spheroids suggest
that the ultraluminous phase of spheroid formation may be visible only
at far infrared wavelengths (Silk \& Wyse 1996; Zepf \& Silk 1996).

Spheroid formation is believed to be triggered by mergers, which
generate a central (within 1~kpc) agglomeration of gas and induce a
central starburst.  Strong winds, whose presence is indirectly
inferred from requirements of early enrichment of intracluster gas,
clear away much of the dust, leaving a reddened nucleus.  Disks form
by subsequent infall of gas, and chemical evolution models strongly
favor an inside-out formation sequence disk of star formation (Wang \&
Silk 1994; Prantzos \& Aubert 1995).  This leads us to the natural
expectation that the early evolution of a spheroid-dominated galaxy
(S0, Sa, Sb) should reveal a dusty, evolved nucleus, surrounded by a
young, vigorously star-forming, relatively low metallicity and so
dust-free, disk.  Only the $K$-band flux can penetrate the dusty
central region; with HST quality imaging in the near-IR, one would
expect in this picture for the $K$ flux to be more centrally
concentrated than the $V$ or $I$ flux for the outliers.  These
observations will become feasible with the NICMOS array on the HST.

Clearly, redshifts would remove much of the ambiguity in this puzzle.
That is a fairly difficult task, particularly if the red oultiers are
absorption-line galaxies.  All of the outliers in this survey are
fainter than $I\simgt26$~mag.  At the moment only the Keck telesope
might be able to obtain spectra for such faint galaxies, and even with
the Keck, spectroscopy for objects this faint will probably have to
wait for adaptive optics to allow very narrow slits to reduce the
foreground sky emission.

\subsection{Brown Dwarfs?}

Finally, do brown dwarfs fit anywhere in the picture?  Recently,
improved atmospheric models for brown dwarfs have emerged (Burrows
1994), as well as actual detections of brown dwarfs (Basri et
al. 1995a; Nakajima et al. 1995).  Models from Burrows (1994), and the
actual datapoint for Basri et al.'s (1995a) detected brown dwarf,
PPl-15 in the Pleiades ($K$ photometry from Basri et al. 1995b; $V$
and $I$ photometry from Stauffer et al. 1994), are plotted in the
color-color diagram (Figure 9).  There are several objects we detect
that are in the vicinity of the locus of points implied by PPl-15 and
the Burrows models.  We survey less than two square arcminutes of
``blank'' sky at rather high Galactic latitudes, so the a priori
chances of intercepting {\it any} star in our lines of sight are
exceedingly slim.

\section{Conclusions}

Our number-magnitude results show that the counts continue to increase
above $K\approx22$~mag, reaching surface densities of
few~$\times~10^5$~degree$^{-2}$.  Fitting the (corrected)
number-magnitude data over the range $18-23$~mag we derive a slope of
d~log~N~/~d~mag~$\approx0.23\pm0.02$.  The optical$-$near-infrared
colors of the extremely faint galaxies show that the majority of
objects are consistent with spirals and irregulars at redshifts around
few tenths or greater.  A few objects have colors consistent with
elliptical galaxies at $z\approx1$.  There is a rather large density
of faint ($K=20-22$) ``red outlier'' galaxies, with blue optical
colors $V-I\simlt2.5$ and red near-infrared colors $I-K\simgt4$.  By
exploring a broad range of Bruzual-Charlot isochrone synthesis
evolutionary models and by a qualitative scenario that involves dust,
several possibilities are identified.  Redshifts, $\sim0.1$~arcsec
quality imaging in the near-infrared, or both will be required to
determine their nature conclusively.

\acknowledgments

We would like to thank Stephane Charlot for his assistance with the
BC93 and the BC95 models, W. Harrison and B. Schaeffer for assistance
with the W. M. Keck observations, and S. D'odorico, M. Tarenghi, and
E. J. Wampler for allowing us to use the NTT observations in advance
of publication.  LAM would particularly like to thank Steve Zepf for
many useful and stimulating conversations, and Mark Dickinson for
frequent de-mystifying and enlightening advice.  The referee Harry
Ferguson provided invaluable feedback on this work.  This work was
supported in part by NSF grant AST92-21540 (MD), by grants from NASA
(JS), and by the Packard Foundation (JRG).

\clearpage

\begin{deluxetable}{lcr}
\tablewidth{10truecm}
\tablecaption{Field Coordinates}
\tablehead{\colhead{Field \#}&\colhead{($\alpha,\delta$)} &
           \colhead{PA ($^{\circ}$)} }
\startdata
Field I  & $10^h 43^m 32^s.1$ $-$00$\arcdeg 01\arcmin 32\farcs0$ (1950)&$0.0$\nl
Field II & $13^h 41^m 52^s.5$ $+$00$\arcdeg 08\arcmin 03\farcs6$ (1950)&$-49.0$\nl
\enddata
\end{deluxetable}

\clearpage

\begin{deluxetable}{ccrrccc}
\tablecaption{Object Catalog for Field I}
\tablehead{\colhead{$\alpha$ (1950)} & \colhead{$\delta$ (1950)} &
           \colhead{$x ('')$}             & \colhead{$y ('')$} &
           \colhead{$V$} & \colhead{$I$} & \colhead{$K$} }
\startdata
10:43:29.74 &$-$00:01:10.85&$   35.4$&$   21.2$&$ 25.7$&$  23.9$&$  22.6$\nl
10:43:29.75 &$-$00:01:36.20&$   35.2$&$   -4.2$&$ 24.3$&$  23.3$&$  22.4$\nl
10:43:29.83 &$-$00:01:09.65&$   34.1$&$   22.4$&$ 25.8$&$  24.3$&$  22.3$\nl
10:43:29.84 &$-$00:01:37.55&$   33.9$&$   -5.6$&$ 26.2$&$  24.8$&$  22.5$\nl
10:43:29.85 &$-$00:01:18.65&$   33.8$&$   13.4$&$  0.0$&$  25.9$&$  21.7$\nl
10:43:29.85 &$-$00:01:59.00&$   33.8$&$  -27.0$&$ 24.4$&$  23.5$&$  21.5$\nl
10:43:29.88 &$-$00:00:59.75&$   33.3$&$   32.2$&$ 27.7$&$   0.0$&$  23.6$\nl
10:43:29.90 &$-$00:01:44.30&$   33.0$&$  -12.3$&$ 23.9$&$  23.3$&$  22.0$\nl
10:43:29.95 &$-$00:01:29.15&$   32.2$&$    2.9$&$ 24.4$&$  22.6$&$  20.6$\nl
10:43:29.98 &$-$00:01:01.85&$   31.8$&$   30.2$&$ 24.5$&$  23.2$&$  22.3$\nl
10:43:30.01 &$-$00:01:08.00&$   31.4$&$   24.0$&$ 24.1$&$  23.2$&$  23.3$\nl
10:43:30.06 &$-$00:01:47.15&$   30.6$&$  -15.2$&$ 25.8$&$  22.9$&$  19.2$\nl
10:43:30.10 &$-$00:01:23.90&$   30.0$&$    8.1$&$ 27.5$&$  24.7$&$  22.5$\nl
10:43:30.10 &$-$00:01:33.20&$   30.0$&$   -1.2$&$ 26.6$&$  25.0$&$  23.1$\nl
10:43:30.24 &$-$00:01:21.35&$   27.9$&$   10.7$&$  0.0$&$  25.6$&$  20.9$\nl
10:43:30.26 &$-$00:01:39.35&$   27.6$&$   -7.4$&$ 24.1$&$  22.8$&$  21.2$\nl
10:43:30.27 &$-$00:01:03.35&$   27.5$&$   28.7$&$ 21.4$&$  19.2$&$  16.7$\nl
10:43:30.34 &$-$00:01:58.10&$   26.4$&$  -26.1$&$ 24.6$&$  23.5$&$  20.8$\nl
10:43:30.36 &$-$00:00:58.70&$   26.1$&$   33.3$&$ 26.0$&$  24.5$&$  21.1$\nl
10:43:30.38 &$-$00:01:31.85&$   25.8$&$    0.2$&$ 26.5$&$  22.8$&$  19.6$\nl
10:43:30.39 &$-$00:01:16.70&$   25.7$&$   15.3$&$ 26.9$&$  25.4$&$  22.9$\nl
10:43:30.51 &$-$00:01:12.80&$   23.9$&$   19.2$&$ 25.5$&$  24.6$&$  22.9$\nl
10:43:30.51 &$-$00:01:28.70&$   23.9$&$    3.3$&$ 27.8$&$  26.2$&$  21.7$\nl
10:43:30.57 &$-$00:01:09.65&$   23.0$&$   22.4$&$ 25.7$&$  25.1$&$  23.2$\nl
10:43:30.57 &$-$00:00:59.75&$   23.0$&$   32.2$&$ 23.5$&$  22.3$&$  21.0$\nl
10:43:30.63 &$-$00:01:30.95&$   22.1$&$    1.1$&$ 24.7$&$  23.8$&$  22.2$\nl
10:43:30.64 &$-$00:02:02.45&$   21.9$&$  -30.5$&$ 26.2$&$  24.5$&$  20.1$\nl
10:43:30.72 &$-$00:01:22.85&$   20.7$&$    9.2$&$ 26.2$&$  24.7$&$  23.5$\nl
10:43:30.77 &$-$00:01:09.50&$   20.0$&$   22.5$&$ 26.8$&$  26.3$&$  23.4$\nl
10:43:30.80 &$-$00:01:21.05&$   19.5$&$   11.0$&$ 23.7$&$  21.0$&$  18.0$\nl
10:43:30.80 &$-$00:01:17.75&$   19.5$&$   14.3$&$ 26.3$&$  26.0$&$  23.8$\nl
10:43:30.81 &$-$00:01:34.55&$   19.4$&$   -2.6$&$ 27.8$&$  26.2$&$  23.7$\nl
10:43:30.88 &$-$00:01:30.20&$   18.3$&$    1.8$&$ 23.8$&$  22.4$&$  20.7$\nl
10:43:30.89 &$-$00:01:25.55&$   18.2$&$    6.5$&$ 27.2$&$  25.7$&$  23.5$\nl
10:43:30.92 &$-$00:01:52.25&$   17.7$&$  -20.2$&$ 27.3$&$  25.8$&$  21.4$\nl
10:43:30.93 &$-$00:01:48.35&$   17.6$&$  -16.4$&$ 27.8$&$  24.7$&$  22.8$\nl
10:43:30.93 &$-$00:01:11.15&$   17.6$&$   20.9$&$ 26.5$&$  24.6$&$  22.6$\nl
10:43:30.93 &$-$00:01:44.75&$   17.6$&$  -12.8$&$ 26.1$&$  25.0$&$  23.1$\nl
10:43:31.01 &$-$00:01:56.30&$   16.4$&$  -24.3$&$ 25.9$&$  24.6$&$  22.1$\nl
10:43:31.04 &$-$00:01:01.10&$   15.9$&$   30.9$&$ 23.7$&$  22.6$&$  20.6$\nl
10:43:31.06 &$-$00:02:05.15&$   15.6$&$  -33.2$&$ 26.8$&$  25.6$&$  22.1$\nl
10:43:31.11 &$-$00:01:22.55&$   14.9$&$    9.5$&$ 24.0$&$  22.6$&$  22.4$\nl
10:43:31.18 &$-$00:01:16.25&$   13.8$&$   15.8$&$ 23.3$&$  21.3$&$  18.8$\nl
10:43:31.20 &$-$00:02:01.40&$   13.5$&$  -29.4$&$  0.0$&$  25.2$&$  23.1$\nl
10:43:31.39 &$-$00:00:56.00&$   10.7$&$   36.0$&$ 26.6$&$  25.4$&$  23.5$\nl
10:43:31.40 &$-$00:01:43.40&$   10.5$&$  -11.4$&$ 25.7$&$  24.9$&$  22.3$\nl
10:43:31.49 &$-$00:01:35.75&$    9.2$&$   -3.8$&$ 24.8$&$  24.1$&$  21.3$\nl
10:43:31.51 &$-$00:01:32.00&$    8.9$&$    0.0$&$ 24.0$&$  22.8$&$  20.7$\nl
10:43:31.53 &$-$00:01:53.00&$    8.6$&$  -21.0$&$ 27.0$&$  26.1$&$  23.6$\nl
10:43:31.56 &$-$00:01:46.25&$    8.1$&$  -14.3$&$ 26.0$&$  24.8$&$  22.5$\nl
10:43:31.58 &$-$00:01:02.30&$    7.8$&$   29.7$&$ 25.8$&$  24.7$&$  22.4$\nl
10:43:31.62 &$-$00:01:55.25&$    7.2$&$  -23.2$&$ 26.8$&$  25.4$&$  24.0$\nl
10:43:31.74 &$-$00:02:01.40&$    5.4$&$  -29.4$&$ 23.6$&$  21.8$&$  20.3$\nl
10:43:31.76 &$-$00:01:26.45&$    5.1$&$    5.6$&$ 26.3$&$  24.6$&$  23.8$\nl
10:43:31.78 &$-$00:00:57.50&$    4.8$&$   34.5$&$  0.0$&$  26.4$&$  22.7$\nl
10:43:31.80 &$-$00:00:59.30&$    4.5$&$   32.7$&$ 27.5$&$  25.5$&$  23.5$\nl
10:43:31.84 &$-$00:01:42.80&$    3.9$&$  -10.8$&$ 25.2$&$  24.3$&$  22.5$\nl
10:43:31.95 &$-$00:01:11.30&$    2.2$&$   20.7$&$ 24.1$&$  22.4$&$  18.9$\nl
10:43:31.96 &$-$00:01:38.75&$    2.1$&$   -6.8$&$ 25.3$&$  24.5$&$  22.7$\nl
10:43:32.00 &$-$00:01:44.60&$    1.5$&$  -12.6$&$ 24.4$&$  23.7$&$  21.4$\nl
10:43:32.04 &$-$00:00:55.55&$    0.9$&$   36.5$&$ 26.7$&$  24.8$&$  21.8$\nl
10:43:32.20 &$-$00:01:36.65&$   -1.5$&$   -4.7$&$ 27.1$&$  26.9$&$  22.5$\nl
10:43:32.23 &$-$00:01:59.45&$   -2.0$&$  -27.5$&$ 25.4$&$  24.3$&$  20.6$\nl
10:43:32.23 &$-$00:00:57.35&$   -2.0$&$   34.7$&$ 26.6$&$  24.5$&$  19.4$\nl
10:43:32.24 &$-$00:01:35.60&$   -2.1$&$   -3.6$&$ 25.9$&$  24.9$&$  22.8$\nl
10:43:32.34 &$-$00:01:22.55&$   -3.6$&$    9.5$&$ 26.2$&$  25.1$&$  23.2$\nl
10:43:32.35 &$-$00:01:16.40&$   -3.8$&$   15.6$&$ 27.4$&$  25.2$&$  23.1$\nl
10:43:32.39 &$-$00:01:40.70&$   -4.4$&$   -8.7$&$ 24.6$&$  22.8$&$  20.6$\nl
10:43:32.41 &$-$00:02:03.20&$   -4.7$&$  -31.2$&$ 27.6$&$  25.5$&$  21.2$\nl
10:43:32.46 &$-$00:01:49.85&$   -5.4$&$  -17.9$&$ 25.7$&$  24.1$&$  21.9$\nl
10:43:32.51 &$-$00:00:59.60&$   -6.2$&$   32.4$&$ 26.5$&$  25.0$&$  23.2$\nl
10:43:32.58 &$-$00:01:24.35&$   -7.2$&$    7.7$&$ 26.2$&$  24.9$&$  23.6$\nl
10:43:32.59 &$-$00:01:28.70&$   -7.4$&$    3.3$&$ 25.5$&$  24.5$&$  21.9$\nl
10:43:32.60 &$-$00:02:02.90&$   -7.5$&$  -30.9$&$ 25.0$&$  22.7$&$  19.2$\nl
10:43:32.74 &$-$00:01:59.45&$   -9.6$&$  -27.5$&$ 26.6$&$  26.0$&$  22.7$\nl
10:43:32.76 &$-$00:01:38.75&$   -9.9$&$   -6.8$&$ 26.3$&$  25.7$&$  22.1$\nl
10:43:32.77 &$-$00:00:55.40&$  -10.1$&$   36.6$&$ 27.5$&$  25.9$&$  24.5$\nl
10:43:32.78 &$-$00:01:56.30&$  -10.2$&$  -24.3$&$ 25.0$&$  23.4$&$  19.9$\nl
10:43:32.82 &$-$00:01:00.80&$  -10.8$&$   31.2$&$  0.0$&$  25.7$&$  23.4$\nl
10:43:32.88 &$-$00:01:16.70&$  -11.7$&$   15.3$&$ 25.0$&$  23.3$&$  20.5$\nl
10:43:32.88 &$-$00:01:38.15&$  -11.7$&$   -6.2$&$ 26.7$&$  25.5$&$  20.4$\nl
10:43:32.89 &$-$00:00:57.20&$  -11.9$&$   34.8$&$ 24.3$&$  23.4$&$  21.2$\nl
10:43:32.91 &$-$00:01:22.40&$  -12.2$&$    9.6$&$ 26.6$&$  25.3$&$  23.5$\nl
10:43:32.91 &$-$00:01:19.40&$  -12.2$&$   12.6$&$ 24.5$&$  23.7$&$  22.3$\nl
10:43:32.98 &$-$00:01:51.80&$  -13.2$&$  -19.8$&$ 25.9$&$  25.0$&$  22.4$\nl
10:43:33.02 &$-$00:01:44.45&$  -13.8$&$  -12.5$&$ 25.7$&$  23.7$&$  20.9$\nl
10:43:33.13 &$-$00:01:10.40&$  -15.5$&$   21.6$&$ 25.5$&$  23.5$&$  21.5$\nl
10:43:33.15 &$-$00:02:01.85&$  -15.8$&$  -29.9$&$ 27.8$&$  26.4$&$  23.1$\nl
10:43:33.16 &$-$00:00:56.00&$  -15.9$&$   36.0$&$ 24.3$&$  22.6$&$  20.3$\nl
10:43:33.18 &$-$00:01:51.35&$  -16.2$&$  -19.4$&$ 24.7$&$  23.5$&$  21.8$\nl
10:43:33.19 &$-$00:01:58.70&$  -16.4$&$  -26.7$&$ 23.7$&$  22.3$&$  20.5$\nl
10:43:33.22 &$-$00:01:12.80&$  -16.8$&$   19.2$&$ 23.0$&$  20.8$&$  18.2$\nl
10:43:33.23 &$-$00:01:35.15&$  -17.0$&$   -3.2$&$ 24.9$&$  23.5$&$  20.3$\nl
10:43:33.24 &$-$00:01:00.05&$  -17.1$&$   32.0$&$ 26.1$&$  25.4$&$  23.7$\nl
10:43:33.33 &$-$00:01:44.60&$  -18.5$&$  -12.6$&$ 24.8$&$  23.7$&$  22.3$\nl
10:43:33.41 &$-$00:01:42.35&$  -19.7$&$  -10.4$&$  0.0$&$  26.3$&$  23.2$\nl
10:43:33.43 &$-$00:01:08.60&$  -20.0$&$   23.4$&$ 22.8$&$  21.3$&$  19.6$\nl
10:43:33.46 &$-$00:00:55.10&$  -20.4$&$   36.9$&$ 26.1$&$  25.3$&$  24.3$\nl
10:43:33.48 &$-$00:02:03.65&$  -20.7$&$  -31.7$&$ 25.7$&$  24.8$&$  20.3$\nl
10:43:33.50 &$-$00:01:24.65&$  -21.0$&$    7.4$&$ 20.5$&$  19.4$&$  17.8$\nl
10:43:33.57 &$-$00:01:40.25&$  -22.1$&$   -8.2$&$ 26.8$&$  25.7$&$  23.4$\nl
10:43:33.65 &$-$00:01:45.95&$  -23.2$&$  -14.0$&$ 25.2$&$  23.7$&$  21.8$\nl
10:43:33.70 &$-$00:01:32.60&$  -24.0$&$   -0.6$&$ 25.7$&$  25.7$&$  22.0$\nl
10:43:33.72 &$-$00:01:17.60&$  -24.3$&$   14.4$&$ 25.9$&$  24.8$&$  22.8$\nl
10:43:33.81 &$-$00:01:21.35&$  -25.7$&$   10.7$&$ 25.3$&$  25.0$&$  22.3$\nl
10:43:33.89 &$-$00:01:01.55&$  -26.9$&$   30.5$&$ 26.6$&$  25.5$&$  24.6$\nl
10:43:33.91 &$-$00:01:36.05&$  -27.2$&$   -4.1$&$ 26.8$&$  25.5$&$  23.6$\nl
10:43:34.01 &$-$00:01:03.80&$  -28.7$&$   28.2$&$ 28.0$&$  23.4$&$  19.5$\nl
10:43:34.04 &$-$00:01:09.65&$  -29.1$&$   22.4$&$ 25.2$&$  23.9$&$  22.2$\nl
10:43:34.06 &$-$00:01:12.20&$  -29.4$&$   19.8$&$ 24.1$&$  22.6$&$  21.0$\nl
10:43:34.13 &$-$00:01:31.10&$  -30.5$&$    0.9$&$ 24.5$&$  23.0$&$  21.8$\nl
10:43:34.30 &$-$00:01:59.15&$  -33.0$&$  -27.2$&$ 26.3$&$  25.0$&$  23.1$\nl
10:43:34.34 &$-$00:01:35.90&$  -33.6$&$   -3.9$&$ 27.4$&$  26.1$&$  23.6$\nl
10:43:34.35 &$-$00:01:20.00&$  -33.8$&$   12.0$&$ 25.8$&$  24.4$&$  19.5$\nl
10:43:34.45 &$-$00:01:48.05&$  -35.2$&$  -16.1$&$ 23.9$&$  22.6$&$  22.4$\nl
10:43:34.48 &$-$00:00:55.40&$  -35.7$&$   36.6$&$ 27.1$&$  24.2$&$  22.5$\nl
\enddata
\end{deluxetable}

\clearpage

\begin{deluxetable}{ccrrccc}
\tablecaption{Object Catalog for Field II}
\tablehead{\colhead{$\alpha$ (1950)} & \colhead{$\delta$ (1950)} &
           \colhead{$x ('')$}             & \colhead{$y ('')$} &
           \colhead{$V$} & \colhead{$I$} & \colhead{$K$} }
\startdata
13:41:49.47 &$+$00:08:00.39&$   27.3$&$  -36.5$&$ 25.1$&$  20.0$&$  19.1$\nl
13:41:49.82 &$+$00:07:59.92&$   23.6$&$  -32.9$&$ 26.5$&$  22.4$&$  23.7$\nl
13:41:49.84 &$+$00:08:09.57&$   30.6$&$  -26.3$&$ 27.9$&$  22.8$&$  21.8$\nl
13:41:49.86 &$+$00:08:15.40&$   34.8$&$  -22.2$&$  0.0$&$  24.1$&$  23.7$\nl
13:41:49.88 &$+$00:08:06.66&$   28.1$&$  -27.8$&$  0.0$&$  23.4$&$  22.5$\nl
13:41:50.21 &$+$00:08:00.89&$   20.4$&$  -27.8$&$  0.0$&$  25.1$&$  22.9$\nl
13:41:50.27 &$+$00:08:04.98&$   23.0$&$  -24.5$&$ 27.3$&$  22.8$&$  22.3$\nl
13:41:50.55 &$+$00:08:17.86&$   29.9$&$  -12.8$&$  0.0$&$  25.2$&$  24.3$\nl
13:41:50.66 &$+$00:08:05.53&$   19.5$&$  -19.7$&$  0.0$&$  24.6$&$  23.4$\nl
13:41:50.66 &$+$00:07:56.38&$   12.6$&$  -25.7$&$ 27.1$&$  22.8$&$  21.7$\nl
13:41:50.71 &$+$00:07:56.44&$   12.2$&$  -25.1$&$  0.0$&$  25.0$&$  22.8$\nl
13:41:50.79 &$+$00:07:56.25&$   11.2$&$  -24.3$&$  0.0$&$  24.7$&$  21.9$\nl
13:41:50.87 &$+$00:07:38.30&$   -3.2$&$  -35.1$&$ 27.8$&$  23.0$&$  22.7$\nl
13:41:50.91 &$+$00:08:18.13&$   26.6$&$   -8.6$&$  0.0$&$  24.7$&$  23.5$\nl
13:41:50.94 &$+$00:07:38.55&$   -3.6$&$  -34.2$&$ 27.8$&$  21.8$&$  21.7$\nl
13:41:51.07 &$+$00:08:10.73&$   19.4$&$  -11.6$&$ 25.9$&$  21.3$&$  20.9$\nl
13:41:51.08 &$+$00:08:28.69&$   32.9$&$    0.3$&$ 26.6$&$  21.9$&$  21.4$\nl
13:41:51.12 &$+$00:07:51.32&$    4.2$&$  -23.7$&$ 27.9$&$  22.9$&$  22.4$\nl
13:41:51.21 &$+$00:08:02.93&$   12.2$&$  -15.2$&$  0.0$&$  26.5$&$  23.6$\nl
13:41:51.22 &$+$00:08:20.78&$   25.5$&$   -3.3$&$ 27.6$&$  23.1$&$  22.6$\nl
13:41:51.31 &$+$00:08:30.47&$   32.0$&$    4.1$&$ 26.7$&$  22.5$&$  21.2$\nl
13:41:51.32 &$+$00:08:26.52&$   28.8$&$    1.7$&$ 27.4$&$  23.0$&$  21.4$\nl
13:41:51.33 &$+$00:08:10.67&$   16.8$&$   -8.7$&$  0.0$&$  25.5$&$  23.7$\nl
13:41:51.37 &$+$00:07:36.39&$   -9.5$&$  -30.8$&$ 25.4$&$  20.7$&$  19.0$\nl
13:41:51.39 &$+$00:07:33.08&$  -12.2$&$  -32.7$&$  0.0$&$  24.2$&$  22.3$\nl
13:41:51.44 &$+$00:07:33.44&$  -12.5$&$  -31.8$&$  0.0$&$  23.8$&$  23.2$\nl
13:41:51.49 &$+$00:07:54.55&$    3.0$&$  -17.4$&$ 26.8$&$  22.4$&$  22.2$\nl
13:41:51.54 &$+$00:08:27.78&$   27.6$&$    5.0$&$  0.0$&$  25.3$&$  23.9$\nl
13:41:51.61 &$+$00:08:31.85&$   30.0$&$    8.4$&$ 22.9$&$  18.6$&$  18.6$\nl
13:41:51.62 &$+$00:07:46.12&$   -4.7$&$  -21.5$&$  0.0$&$  24.6$&$  23.9$\nl
13:41:51.67 &$+$00:07:36.18&$  -12.6$&$  -27.5$&$  0.0$&$  22.1$&$  19.8$\nl
13:41:51.68 &$+$00:08:12.74&$   14.9$&$   -3.3$&$ 27.3$&$  23.1$&$  22.3$\nl
13:41:51.69 &$+$00:07:58.91&$    4.4$&$  -12.3$&$ 27.1$&$  22.3$&$  21.8$\nl
13:41:51.72 &$+$00:07:57.08&$    2.7$&$  -13.2$&$ 27.4$&$  21.0$&$  18.7$\nl
13:41:51.75 &$+$00:08:39.04&$   34.1$&$   14.7$&$  0.0$&$  24.6$&$  23.2$\nl
13:41:51.80 &$+$00:08:15.18&$   15.6$&$   -0.5$&$  0.0$&$  24.8$&$  23.6$\nl
13:41:52.00 &$+$00:07:44.87&$   -9.3$&$  -18.0$&$  0.0$&$  22.3$&$  20.1$\nl
13:41:52.02 &$+$00:08:29.62&$   24.3$&$   11.6$&$  0.0$&$  24.3$&$  23.9$\nl
13:41:52.02 &$+$00:08:35.78&$   29.0$&$   15.6$&$ 27.0$&$  23.1$&$  22.1$\nl
13:41:52.06 &$+$00:07:26.20&$  -24.0$&$  -29.6$&$ 27.8$&$  23.4$&$  22.0$\nl
13:41:52.08 &$+$00:08:10.95&$    9.6$&$    0.0$&$ 26.2$&$  20.8$&$  19.3$\nl
13:41:52.11 &$+$00:08:26.01&$   20.7$&$   10.2$&$  0.0$&$  24.2$&$  22.8$\nl
13:41:52.24 &$+$00:07:48.35&$   -9.0$&$  -13.1$&$ 27.9$&$  22.8$&$  22.4$\nl
13:41:52.27 &$+$00:08:26.37&$   19.4$&$   12.3$&$  0.0$&$  26.3$&$  22.7$\nl
13:41:52.29 &$+$00:08:06.26&$    4.1$&$   -0.8$&$  0.0$&$  24.1$&$  24.0$\nl
13:41:52.35 &$+$00:08:36.29&$   26.1$&$   19.7$&$ 25.7$&$  21.4$&$  20.9$\nl
13:41:52.41 &$+$00:07:47.63&$  -11.2$&$  -11.6$&$  0.0$&$  24.5$&$  24.8$\nl
13:41:52.42 &$+$00:08:40.36&$   28.5$&$   23.1$&$  0.0$&$  24.1$&$  23.6$\nl
13:41:52.43 &$+$00:08:26.63&$   18.0$&$   14.3$&$  0.0$&$  25.9$&$  24.1$\nl
13:41:52.47 &$+$00:07:53.95&$   -7.1$&$   -6.8$&$  0.0$&$  23.4$&$  23.9$\nl
13:41:52.55 &$+$00:08:07.48&$    2.4$&$    3.0$&$  0.0$&$  24.5$&$  24.3$\nl
13:41:52.65 &$+$00:08:28.85&$   17.6$&$   18.2$&$  0.0$&$  25.3$&$  24.3$\nl
13:41:52.65 &$+$00:07:23.14&$  -32.1$&$  -24.9$&$ 27.9$&$  22.9$&$  22.4$\nl
13:41:52.65 &$+$00:08:48.62&$   32.4$&$   31.2$&$ 27.0$&$  21.3$&$  20.0$\nl
13:41:52.67 &$+$00:08:35.74&$   22.5$&$   23.0$&$ 27.0$&$  22.8$&$  22.4$\nl
13:41:52.68 &$+$00:08:02.87&$   -2.4$&$    1.5$&$ 26.6$&$  22.0$&$  22.0$\nl
13:41:52.73 &$+$00:08:32.91&$   19.8$&$   21.8$&$  0.0$&$  25.5$&$  23.8$\nl
13:41:52.74 &$+$00:08:26.31&$   14.7$&$   17.6$&$  0.0$&$  23.8$&$  22.9$\nl
13:41:52.77 &$+$00:08:29.68&$   17.0$&$   20.1$&$ 25.6$&$  20.6$&$  20.0$\nl
13:41:52.78 &$+$00:08:42.96&$   26.9$&$   29.0$&$ 27.4$&$  21.2$&$  19.6$\nl
13:41:52.83 &$+$00:08:03.46&$   -3.5$&$    3.6$&$  0.0$&$  24.8$&$  24.3$\nl
13:41:52.84 &$+$00:07:52.18&$  -12.0$&$   -3.8$&$ 27.3$&$  22.4$&$  22.5$\nl
13:41:52.89 &$+$00:08:06.59&$   -1.7$&$    6.3$&$  0.0$&$  25.9$&$  24.9$\nl
13:41:52.89 &$+$00:08:36.46&$   20.9$&$   26.0$&$ 27.9$&$  23.5$&$  22.4$\nl
13:41:52.94 &$+$00:07:21.77&$  -36.0$&$  -22.5$&$  0.0$&$  23.8$&$  24.4$\nl
13:41:52.95 &$+$00:08:22.47&$    9.8$&$   17.4$&$ 28.0$&$  23.5$&$  23.1$\nl
13:41:52.99 &$+$00:08:08.50&$   -1.2$&$    8.7$&$  0.0$&$  23.4$&$  23.2$\nl
13:41:53.02 &$+$00:08:40.26&$   22.5$&$   29.9$&$ 26.4$&$  21.8$&$  21.3$\nl
13:41:53.03 &$+$00:08:36.85&$   19.8$&$   27.8$&$  0.0$&$  22.9$&$  22.5$\nl
13:41:53.09 &$+$00:08:35.83&$   18.5$&$   27.8$&$  0.0$&$  25.3$&$  23.9$\nl
13:41:53.15 &$+$00:08:23.86&$    8.9$&$   20.6$&$ 26.3$&$  21.9$&$  20.9$\nl
13:41:53.16 &$+$00:07:31.75&$  -30.6$&$  -13.5$&$  0.0$&$  24.2$&$  23.8$\nl
13:41:53.21 &$+$00:07:52.74&$  -15.3$&$    0.9$&$  0.0$&$  24.4$&$  22.7$\nl
13:41:53.23 &$+$00:08:15.05&$    1.4$&$   15.8$&$  0.0$&$  24.5$&$  24.4$\nl
13:41:53.25 &$+$00:08:26.20&$    9.6$&$   23.2$&$  0.0$&$  24.9$&$  24.3$\nl
13:41:53.25 &$+$00:07:42.71&$  -23.2$&$   -5.2$&$ 26.8$&$  21.9$&$  21.1$\nl
13:41:53.32 &$+$00:07:45.40&$  -21.9$&$   -2.7$&$ 24.8$&$  19.3$&$  18.2$\nl
13:41:53.39 &$+$00:08:12.12&$   -2.4$&$   15.6$&$  0.0$&$  24.1$&$  24.0$\nl
13:41:53.43 &$+$00:07:37.52&$  -29.0$&$   -6.6$&$  0.0$&$  23.6$&$  23.4$\nl
13:41:53.45 &$+$00:08:12.28&$   -2.9$&$   16.4$&$  0.0$&$  25.9$&$  24.0$\nl
13:41:53.48 &$+$00:07:56.40&$  -15.2$&$    6.3$&$  0.0$&$  22.1$&$  19.7$\nl
13:41:53.56 &$+$00:07:44.73&$  -24.8$&$   -0.5$&$  0.0$&$  25.0$&$  24.2$\nl
13:41:53.72 &$+$00:07:35.63&$  -33.2$&$   -4.7$&$ 24.8$&$  19.6$&$  20.0$\nl
13:41:53.72 &$+$00:07:51.04&$  -21.6$&$    5.6$&$ 25.4$&$  19.6$&$  18.6$\nl
13:41:53.77 &$+$00:07:46.85&$  -25.2$&$    3.3$&$ 27.5$&$  22.6$&$  21.3$\nl
13:41:53.79 &$+$00:08:31.80&$    8.6$&$   33.0$&$ 23.7$&$  18.8$&$  18.2$\nl
13:41:53.80 &$+$00:08:10.31&$   -7.8$&$   19.1$&$ 27.9$&$  24.6$&$  22.1$\nl
13:41:53.84 &$+$00:08:24.19&$    2.2$&$   28.7$&$ 27.9$&$  23.6$&$  24.0$\nl
13:41:53.95 &$+$00:08:05.27&$  -13.1$&$   17.4$&$  0.0$&$  25.0$&$  24.0$\nl
13:41:54.02 &$+$00:07:51.69&$  -24.0$&$    9.3$&$ 28.0$&$  24.0$&$  22.5$\nl
13:41:54.05 &$+$00:08:15.15&$   -6.6$&$   25.1$&$ 26.6$&$  22.1$&$  21.8$\nl
13:41:54.07 &$+$00:08:20.98&$   -2.4$&$   29.1$&$ 24.8$&$  19.3$&$  18.9$\nl
13:41:54.08 &$+$00:08:07.15&$  -12.9$&$   20.1$&$ 24.8$&$  20.7$&$  20.9$\nl
13:41:54.16 &$+$00:08:03.03&$  -16.8$&$   18.3$&$ 25.1$&$  20.2$&$  20.0$\nl
13:41:54.20 &$+$00:07:59.47&$  -20.0$&$   16.5$&$  0.0$&$  24.0$&$  23.2$\nl
13:41:54.21 &$+$00:08:18.39&$   -5.7$&$   29.0$&$ 25.0$&$  19.3$&$  18.3$\nl
13:41:54.27 &$+$00:07:55.99&$  -23.2$&$   15.0$&$ 26.2$&$  21.7$&$  22.0$\nl
13:41:54.34 &$+$00:08:14.32&$  -10.1$&$   27.8$&$  0.0$&$  24.8$&$  24.0$\nl
13:41:54.41 &$+$00:08:21.89&$   -5.1$&$   33.6$&$ 25.2$&$  19.4$&$  18.9$\nl
13:41:54.42 &$+$00:08:17.63&$   -8.4$&$   30.9$&$  0.0$&$  23.7$&$  24.3$\nl
13:41:54.45 &$+$00:07:44.54&$  -33.6$&$    9.5$&$ 26.1$&$  21.4$&$  19.8$\nl
13:41:54.47 &$+$00:07:53.23&$  -27.3$&$   15.5$&$ 28.0$&$  24.0$&$  23.8$\nl
13:41:54.60 &$+$00:08:16.81&$  -10.8$&$   32.4$&$  0.0$&$  26.4$&$  23.1$\nl
13:41:54.63 &$+$00:08:07.42&$  -18.2$&$   26.6$&$ 27.5$&$  23.3$&$  24.3$\nl
13:41:54.72 &$+$00:08:03.61&$  -21.9$&$   25.1$&$ 24.7$&$  19.4$&$  18.2$\nl
13:41:54.93 &$+$00:07:51.69&$  -33.0$&$   19.7$&$ 27.2$&$  23.0$&$  22.1$\nl
13:41:54.97 &$+$00:08:00.48&$  -26.7$&$   25.8$&$  0.0$&$  24.6$&$  25.0$\nl
13:41:55.06 &$+$00:07:49.96&$  -35.6$&$   20.0$&$  0.0$&$  25.7$&$  24.0$\nl
13:41:55.24 &$+$00:08:02.52&$  -27.9$&$   30.3$&$ 27.5$&$  21.3$&$  20.8$\nl
13:41:55.29 &$+$00:08:00.56&$  -29.9$&$   29.6$&$  0.0$&$  24.5$&$  24.5$\nl
13:41:55.54 &$+$00:07:58.60&$  -33.8$&$   31.1$&$  0.0$&$  23.7$&$  23.6$\nl
\enddata
\end{deluxetable}

\clearpage
\begin{deluxetable}{cccccccc}
\tablecolumns{8}
\tablecaption{Completeness Matrix for 40/60 Ellipticals/Spirals}
\tablehead{
\colhead{} & \multicolumn{7}{c}{Input Magnitudes $m_{input}$} \\
\cline{2-8} \\
\colhead{$m_{recovered}$}  & \colhead{\bf 18} & \colhead{\bf 19} & 
\colhead{\bf 20} & \colhead{\bf 21} & \colhead{\bf 22} & 
\colhead{\bf 23} & \colhead{\bf 24} }
\startdata
{\bf 18}&0.813 & 0.081 & 0.000 & 0.000 & 0.000 & 0.000 & 0.000\nl
{\bf 19}&0.048 & 0.779 & 0.094 & 0.000 & 0.000 & 0.000 & 0.000\nl
{\bf 20}&0.000 & 0.079 & 0.743 & 0.064 & 0.000 & 0.000 & 0.000\nl
{\bf 21}&0.000 & 0.000 & 0.086 & 0.716 & 0.063 & 0.000 & 0.000\nl
{\bf 22}&0.000 & 0.000 & 0.000 & 0.117 & 0.555 & 0.078 & 0.006\nl
{\bf 23}&0.000 & 0.000 & 0.000 & 0.005 & 0.136 & 0.259 & 0.057\nl
{\bf 24}&0.000 & 0.000 & 0.000 & 0.000 & 0.010 & 0.046 & 0.030\nl
\enddata
\end{deluxetable}

\clearpage

\begin{deluxetable}{cccccc}
\tablecaption{Galaxy counts for both Fields}
\tablehead{
\colhead{$K$ mag} & \colhead{$N$} & \colhead{$N_{corr}$} &
\colhead{log~$N$~mag~$^{-1}$~deg$^{-2}$} &
\colhead{log~$N_{-1\sigma}$~mag~$^{-1}$~deg$^{-2}$} &
\colhead{log~$N_{+1\sigma}$~mag~$^{-1}$~deg$^{-2}$} }
\startdata
{\bf 18} & 10 & 11.6 & 4.239 & 4.077 & 4.393\nl
{\bf 19} & 11 & 11.7 & 4.243 & 4.090 & 4.389\nl
{\bf 20} & 26 & 28.2 & 4.624 & 4.401 & 4.716\nl
{\bf 21} & 35 & 38.8 & 4.763 & 4.683 & 4.842\nl
{\bf 22} & 57 & 74.7 & 5.047 & 4.999 & 5.117\nl
{\bf 23} & 48 & 125  & 5.272 & 5.205 & 5.339\nl
{\bf 24} & 29 & 311  & 5.667 & 5.579 & 5.755\nl
\enddata
\end{deluxetable}

\clearpage

{\centerline {\bf Figure Captions}}

\noindent Figure 1 - Contour plot for Field I.  The lowest contour
corresponds to $3\sigma$ above the effective sky (as described in the
text); the remaining contours are plotted in $3\sigma$ intervals.  The
orientation of the Field is as shown.

\noindent Figure 2 - Contour plot for Field II.  Same setup as in Figure 1.

\noindent Figure 3 (Plate 1) - ``True-color'' picture of Field I.
The $K$, $I$, and $V$ images were used for each of the red, green, and
blue color planes.

\noindent Figure 4 (Plate 2) - ``True-color'' picture of Field II.  Same
setup as in Figure 3 (Plate 1).

\noindent Figure 5 - Fixed ($3.0$~arcsec) aperture magnitudes versus metric
($20h^{-1}$~kpc) aperture magnitudes as a function of redshift for
$q_0=0.1$ (solid line), and for $q_0=0.5$ (long-dashed line).  A
galaxy surface brightness profile of $L(<r)\propto r^{0.4}$ (see
Section 3.1) is used.  This Figure illustrates that the offset between
fixed aperture magnitudes and metric apertures is between $0.3$~mag and
$0.1$~mag over $0.4<z<1.0$.

\noindent Figure 6 - The output from one of the Monte-Carlo simulations for
Field I as described in Section 3.2.  By plotting the recovered
magnitude minus the actual total magnitude that was input against the
input magnitude, we recover both the constant-magnitude offset and the
recovery efficiency.  These corrections are then applied to the galaxy
counts, to recover the actual number-magnitude relation shown in
Figure 7 and discussed in Section 3.4.

\noindent Figure 7 - Number-magnitude plot for the completeness-corrected
data.  The error bars are based on the $1\sigma$ Poisson uncertainties
(Gehrels 1986) for the {\it raw} counts.  The insert describes the
efficiency with which objects are recovered at {\it any} magnitude, as
deduced from the simulations.  The efficiency at brighter magnitudes
is not quite 1.0 because some small amount of object overlap occurs at
the 2-3 percent level in the simulations.  The Djorgovski et
al. (1995) data, which are the deepest to date, appear to report
consistently smaller numbers than those reported here.  The Hawaii
Deep Survey points are from Gardner, Cowie, \& Wainscoat (1993).  The
models are based on Yoshii \& Peterson (1995).  They do not explore
luminosity or density evolution of galaxies.  ``Model 1'' uses
$(\Omega_0,
\lambda_0, h_0)$ = (1.0, 0.0, 0.5), and ``model 2'' has (0.2, 0.0,
0.6).  The low-$\Omega$ models are better fits for the data, though a
fit to counts from a single band does not provide a realistic
constraint without simultaneously addressing the counts in other bands
and the redshift distributions.

\noindent Figure 8 - Color-magnitude diagram for all data.  The magnitudes
used here are measured using the isophotes defined in the $VIK$ image.
The error bars are the $1\sigma$ photometric uncertainties.  Stellar
population evolution models are overlaid (BC93) for several
evolutionary histories, with redshifts labeled at intervals.  All
models are normalized to a total mass of $M=10^{11}M_{\odot}$.

\noindent Figure 9 - Color-color diagram for all data.  The magnitudes used
here are measured using the isophotes defined in the $VIK$ image.  A
typical error bar is shown.  The limits of objects which did not have
a reliable detection in one of the bandpasses are shown using the
appropriate photometric limit.  Stellar population models which assume
a formation epoch at $z_{form}=5$ are overlaid (BC93) for several
evolutionary histories.  The heavy solid line depicts an unevolving
elliptical galaxy (i.e. pure K-correction).  The light-solid line and
the short-dashed line are both exponentially-declining star formation
rate (SFR) galaxies, with timescales of $0.1$~Gyr and $5.0$~Gyr
respectively.  The numbers near the squares and rectangles
on these curves denote redshifts.
The $K\approx20$ galaxies that are very red in both
$V-I$ and $I-K$, and the several (``red outlier'') galaxies with
$K\simgt20$ galaxies with $I-K\simgt4$ and $V-I\approx0-2$ are
discussed at length in the text.  The dotted lines are the stellar
sequences (Johnson 1966), and the brown dwarf PPl 15 and several brown
dwarf models are depicted as well (Burrows 1994).

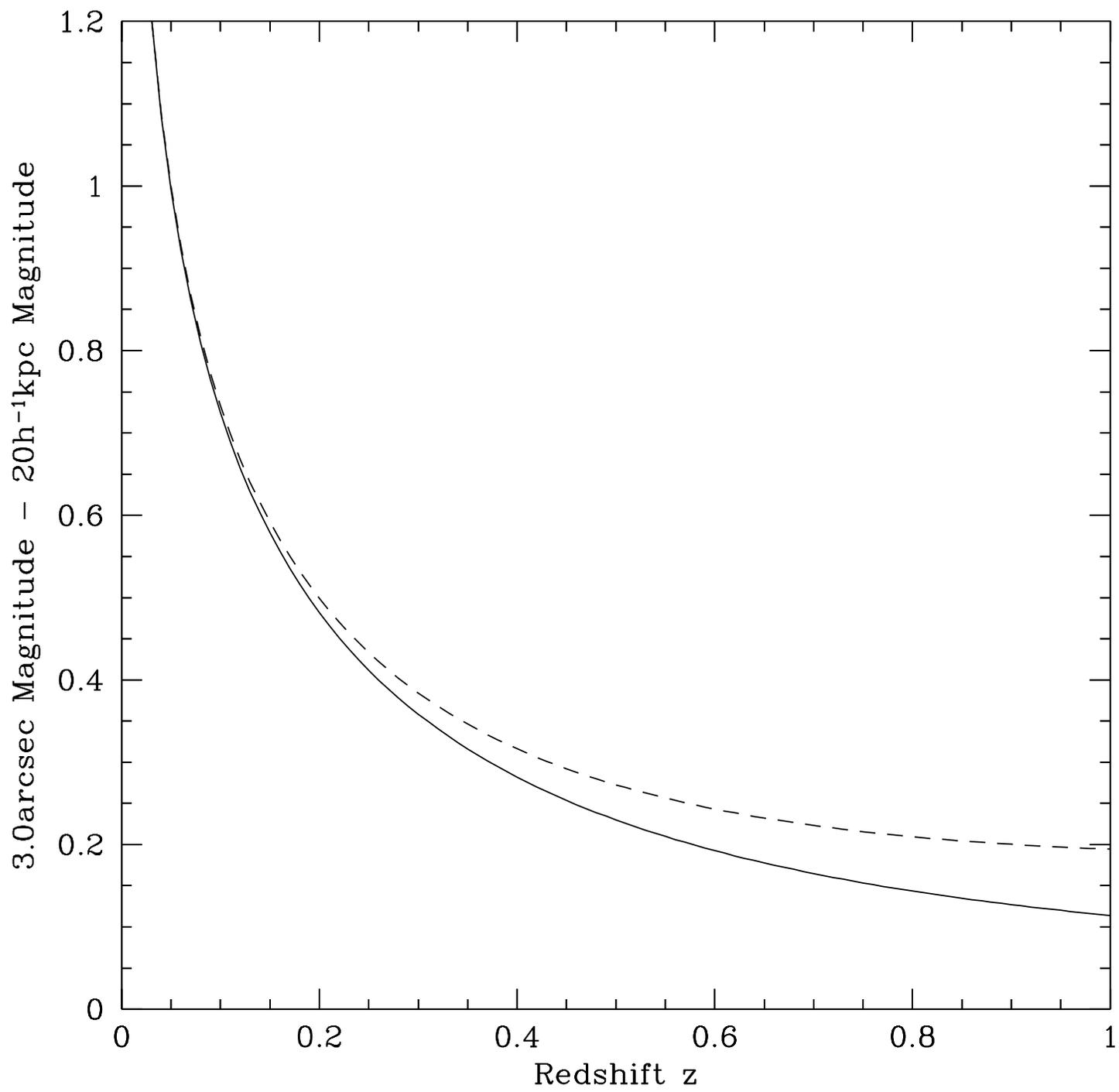
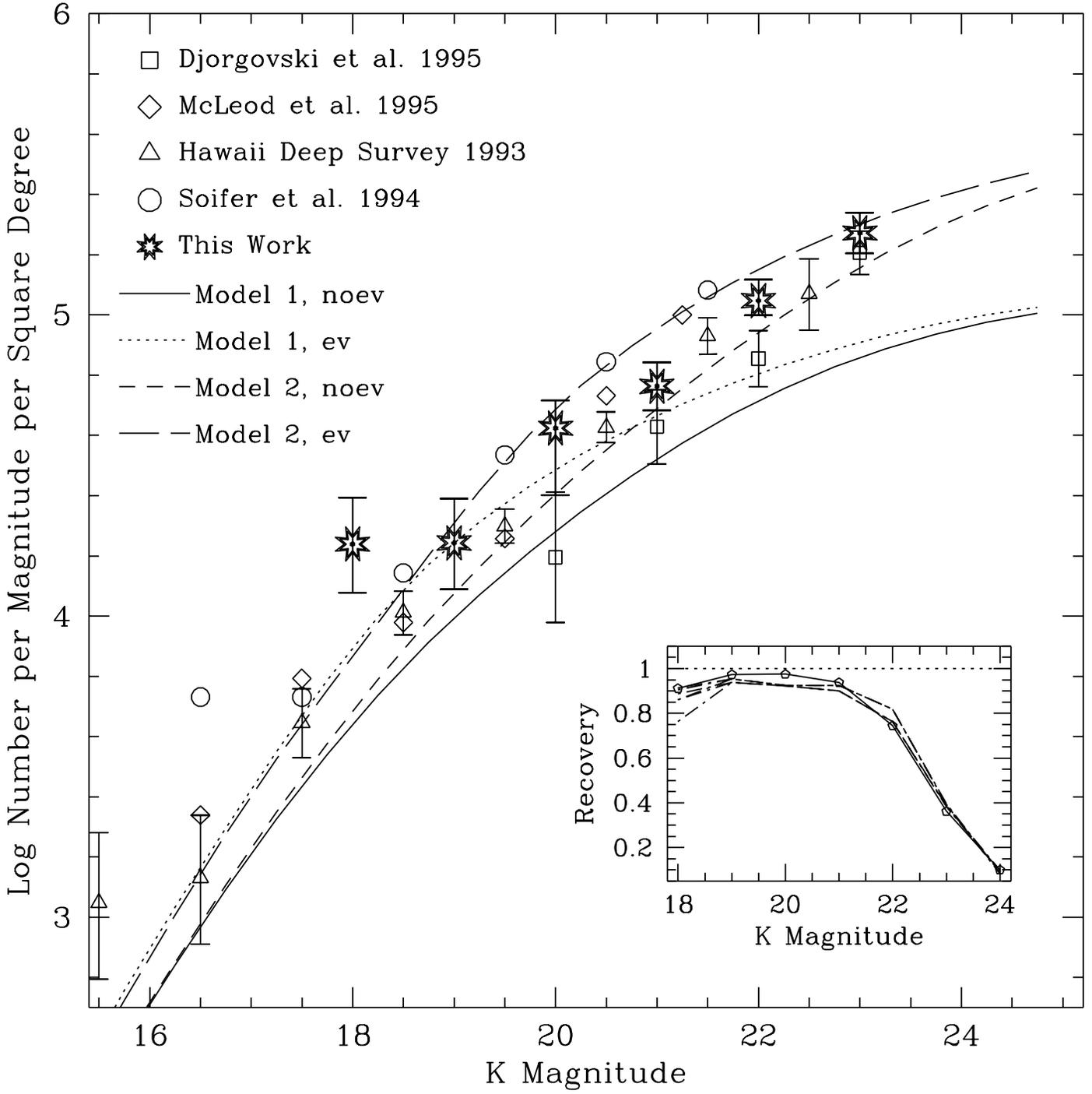
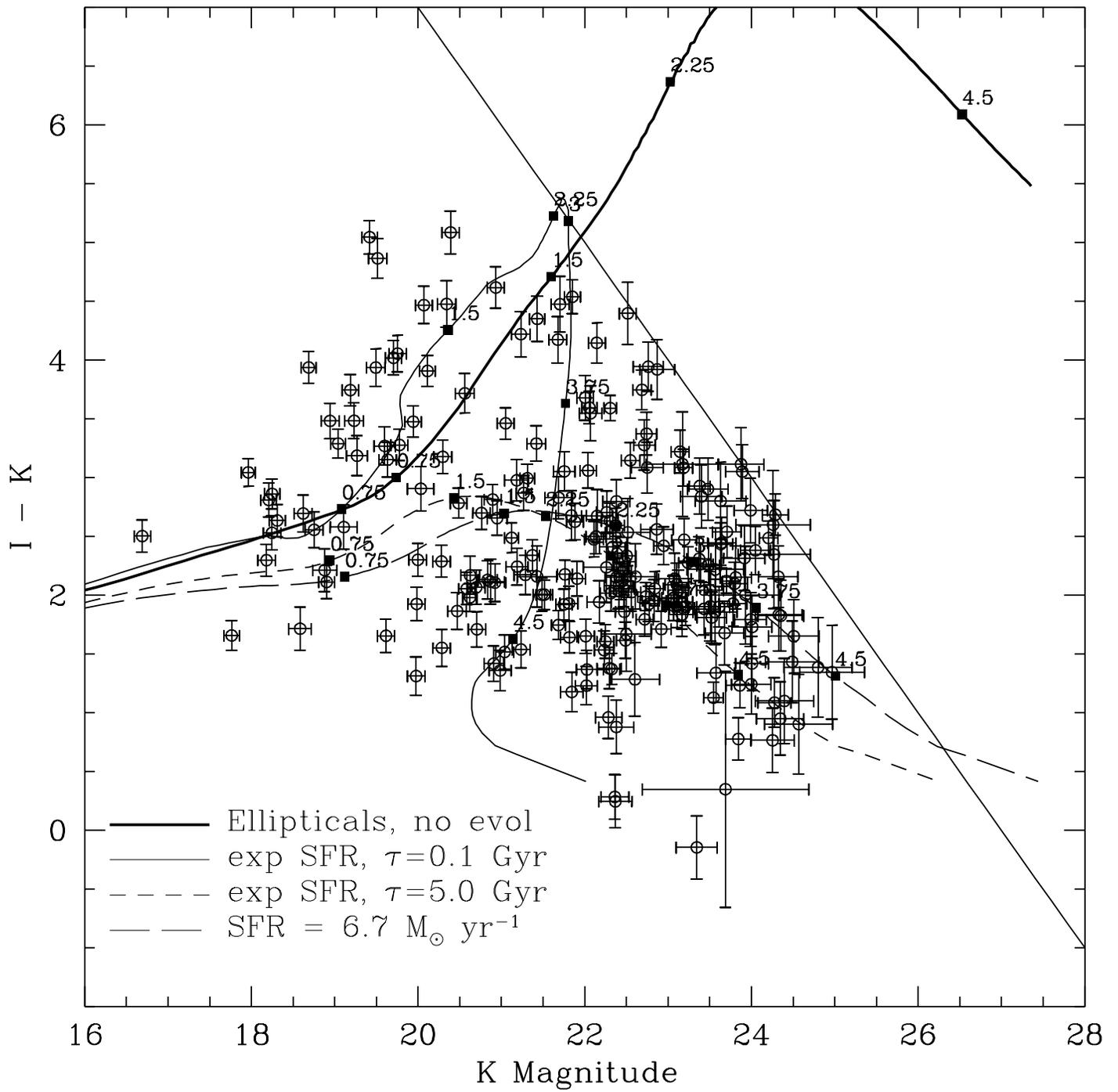
\end{document}